\documentclass[journal,10pt]{IEEEtran}
\usepackage{float}
\usepackage{xcolor}
\usepackage{blindtext}
\usepackage{graphicx}
\usepackage{amsmath}
\usepackage[ruled,vlined]{algorithm2e} 
\usepackage{listings}
\usepackage{siunitx}
\usepackage{balance}
\usepackage[noadjust]{cite} % make [1]-[3] insted of [1],[2],[3]
\usepackage{subcaption}
\usepackage{makecell} % multi-row cell in table
\usepackage{amsfonts} % for mathbb

\SetCommentSty{mycommfont}

\allowdisplaybreaks

\begin{document}
\title{A Hybrid DC Fault Primary Protection Algorithm for Multi-Terminal HVdc Systems}

\author{Jingfan Sun,~\IEEEmembership{Student Member,~IEEE,}
        Suman Debnath,~\IEEEmembership{Senior Member,~IEEE,}\\
        Matthieu Bloch,~\IEEEmembership{Senior Member,~IEEE,}
        Maryam Saeedifard,~\IEEEmembership{Senior Member,~IEEE}
        % Phani R V Marthi,~\IEEEmembership{Student Member,~IEEE,}% <-this % stops a space
\thanks{Jingfan Sun, Matthieu Bloch and Maryam Saeedifard are with the School of Electrical and Computer Engineering at Georgia Institute of Technology, Atlanta, GA, 30332-0250, USA (e-mails: jingfan@gatech.edu; matthieu.bloch@ece.gatech.edu; maryam@ece.gatech.edu).

Suman Debnath is with Oak Ridge National Laboratory, Knoxville, TN 37932, USA (email: debnaths@ornl.gov).}% <-this % stops a space

}

\maketitle

\begin{abstract}

Protection against dc faults is one of the main technical hurdles faced when operating converter-based HVdc systems. Protection becomes even more challenging for multi-terminal dc (MTdc) systems with more than two terminals/converter stations. In this paper, a hybrid primary fault detection algorithm for MTdc systems is proposed to detect a broad range of failures. Sensor measurements, i.e., line currents and dc reactor voltages measured at local terminals, are first processed by a top-level context clustering algorithm. For each cluster, the best fault detector is selected among a detector pool according to a rule resulting from a learning algorithm. The detector pool consists of several existing detection algorithms, each performing differently across fault scenarios. The proposed hybrid primary detection algorithm: i) detects all major fault types including pole-to-pole (P2P), pole-to-ground (P2G), and external dc fault; ii) provides a wide detection region covering faults with various fault locations and impedances; iii) is more robust to noisy sensor measurements compared to the existing methods. Performance and effectiveness of the proposed algorithm are evaluated and verified based on time-domain simulations in the PSCAD/EMTDC software environment. The results confirm satisfactory operation, accuracy, and detection speed of the proposed algorithm under various fault scenarios.

%\boldmath
% \blindtext[1]
\end{abstract}

\begin{IEEEkeywords}
% IEEEtran, journal, \LaTeX, paper, template.
% Multi-terminal HVdc systems, dc-side fault, Backup Protection
Multi-terminal HVdc systems, Dc-side faults, Fault detection
\end{IEEEkeywords}

% \fi
\IEEEpeerreviewmaketitle

\section{Introduction}

\IEEEPARstart{H}{igh} Voltage dc (HVdc) transmission is a mature technology with many installations around the world \cite{1,2,CIGRE1}. The attractive features of Voltage-Sourced Converters (VSCs) along with their recent significant breakthroughs have made the HVdc technology even more promising. 
% The VSCs provide enhanced reliability with reduced cost and power losses. 
Concomitantly, significant changes in generation, transmission, and loads have emerged. These changes include integrating and tapping renewable energy in remote areas, relocating or retiring older conventional and/or nuclear power plants, increasing transmission capacity, and urbanization \cite{2}. These new trends have called for VSC-based Multi-Terminal dc (MTdc) systems, which when embedded inside the ac grid, can enhance stability, reliability, and efficiency of the present power grid \cite{CIGRE1}. 

% talk about the background of EMT simulation, why EMT? what EMT models we have for MMC?
The strategic importance of MTdc systems is evidenced by the number of worldwide projects currently in advanced planning stage, e.g., European “Supergrids” and the Baltic Sea project, along with a few projects in China \cite{1,2}. In the US, the cross-continental dc/ac grid is expected to enable the connection of three primary interconnections \cite{ORNL_report}, i.e., the Western Interconnection (WI), Eastern Interconnection (EI), and Electric Reliability Council of Texas (ERCOT). 

Amid the optimism surrounding the benefits of MTdc systems, their protection against dc-side faults remains one of the major technical challenges \cite{1,2}. 
% Dc fault clearance in MTDC systems, without causing a large loss-of-infeed, have been investigated by either (i) using dc circuit breakers to isolate only the faulty cable/line while continuing to operate the rest of the DC grid as usual \cite{breaker1,breaker2,breaker3,breaker4}, or (ii) using VSC topologies with dc fault current interruption capability \cite{Hybrid_MMC}. As the dc circuit breaker technology becomes more mature and economically viable, the former seems to be more promising. Nevertheless, 
Protection based on dc circuit breakers requires fast, accurate, and selective relaying algorithms to detect faults in the dc systems \cite{breaker1,breaker3,breaker4}. In HVdc systems, compared to their ac counterpart, dc fault detection is far more challenging because of the rapid rise of fault current and low impedance nature of the dc cables/lines \cite{1,breaker1}. The protection philosophy of the MTdc systems, nevertheless, is similar to ac protection in the sense that both primary and backup protection functionalities are required \cite{1,backup}. The primary protection, which is designed to work under normal operation, should respond to any types of faults in a fast and reliable manner.

The study of primary fault detection in MTdc systems has attracted much interest over the past few years, and various algorithms have been proposed. To improve fault detection speed, as well as reliability of the protection system, methods based on only local measurements without relying on any communication across multiple MTdc terminals have attracted significant interest \cite{detection1,detection7,detection9,detection8,detection2,detection3,detection4,detection6}. These algorithms are summarized in Table \ref{literature} and can be categorized by the type of sensor measurements, i.e., currents and/or voltages measured at one end of the transmission line \cite{detection1,detection7,detection9,detection8,detection3}, and voltages measured across the dc current limiting reactors \cite{detection4,detection2,detection6}. 

% https://tex.stackexchange.com/questions/149603/two-figures-in-one-page-of-two-columns-style
\begin{table*}[ht]
\centering
\caption{The existing primary fault detection algorithms for MTdc systems}
% \subfloat{
\begin{tabular}{ccccccccc}
\hline
Paper & Year & Method & Sensor Measurements & Covered Fault Types & Fault Characteristics
 \\\hline
\cite{detection1} & 2020 & threshold, transient average current & line currents & P2P, external & different locations, impedances\\
\cite{detection7} & 2017 & threshold, transient current & line currents & P2G, P2P, external & breaker tripping, line \& converter outage\\
\cite{detection9} & 2012 &  Fourier transform & line currents & P2G, external & different locations, impedances\\
% \cite{detection5} & 2017 & current differential & multiple line currents & between terminals & P2G, P2P, external dc & locations\\
\cite{backup} & 2019 & QCD & line voltages & P2G, P2P & different locations, impedances, power flow, noise\\
\cite{detection8} & 2016 &  ROCOV & line voltages & P2G, P2P, external & different locations\\
\cite{detection3} & 2017 & ROTV \& transient voltage & line voltages & P2G, P2P, external & different locations, impedances\\
\cite{detection4} & 2017 & ROCOV & dc reactor voltages & P2G, P2P & different locations, impedances, power reversal\\
\cite{detection2} & 2018 & threshold, absolute value of voltage & dc reactor voltages & P2G, P2P & different locations, impedances\\
\cite{detection6} & 2019 & travelling wave & dc reactor voltages & P2G, P2P & sampling frequencies, reactor sizes\\
\hline
\multicolumn{2}{c}{Proposed Method} & \makecell{hybrid of threshold, \\ ROCOV, and QCD-based} & \makecell{line currents \\ line voltages} & \makecell{P2G, P2P, external} & \makecell{different locations, impedances, \\ reactor sizes, noisy sensor measurements}\\
\hline
\end{tabular}
% }
\label{literature}
\end{table*}

\begin{figure*}[ht]
\centering
% \subfloat{
\includegraphics[width=0.6\textwidth]{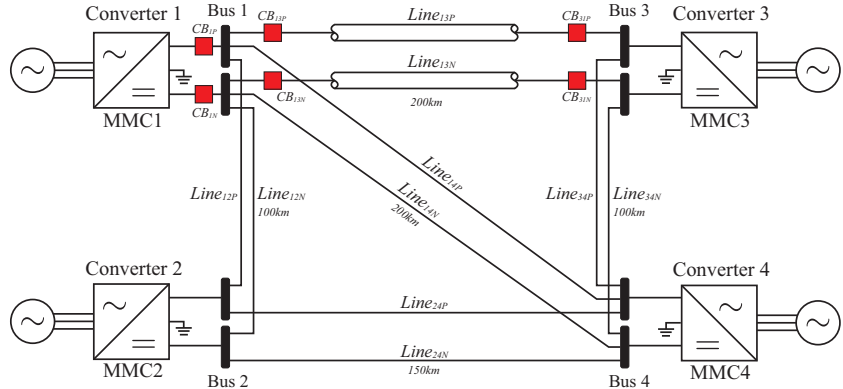}
% }
\caption{Layout of the four-terminal MTdc test system \cite{test_system1}.}
\label{test_system}
\vspace{-1\baselineskip}
\end{figure*}

Upon occurrence of a dc fault, the dc capacitors at the terminals on both ends of the faulty transmission line start to discharge \cite{breaker_selection}. Following this discharge, line current increases rapidly from its nominal value. The fast rising rate of the transient current is used as an indicator of a dc fault in \cite{detection7}. However, this method may fail to detect high impedance faults, during which the rate of rise of fault current is not as high as the low impedance cases. The average value of transient current over a fixed window is used in \cite{detection1} but at the expense of sacrificing the detection speed. Line currents are processed differently in \cite{detection9} by extracting the transient harmonic from the signal through a discrete Fourier transform. This information is used for fault detection and identification of the fault types. Since this method relies on a single frequency component of the fault current, it is vulnerable to signal noise. In addition to the rising fault current, voltages at faulty terminals also drop abruptly upon the arrival of travelling waves from the fault location on the transmission line. This abrupt change in voltage measurements is monitored using the Quickest Change Detection (QCD)-based algorithm \cite{backup}, which offers improved robustness to noise at the expense of slower detection speed. The rate of change of voltage (ROCOV) measured at the transmission line side of the current limiting reactor is used to determine the fault \cite{detection8}. This method is sensitive to fault impedance and can be affected by high-frequency noise, as well. In \cite{detection3}, the transient voltages measured at both sides of the current limiting reactor are collected and combined into the ratio of the transient voltages (ROTV). This primary fault detection method is designed to work with a backup method, which requires the communication of the ROTV from both ends of the line, hence reducing the detection speed.

The other category of primary fault detection algorithms takes advantage of the voltage signal measured across the current limiting reactor connected in series with the dc circuit breaker. The absolute value of this signal is used in \cite{detection2} for primary detection. In \cite{detection4}, the rate of change of dc reactor voltage is monitored. This is essentially equivalent to monitoring the second derivative of the dc fault current. However, a high sampling frequency is required for such measurement, which is neither realistic nor cost-effective in real applications. Another usage of dc reactor voltage is proposed in \cite{detection6}, where the authors extract the positive-sequence voltage from travelling waves to detect the fault. Relying on transient travelling wave signals, this method is vulnerable to noisy sensor measurements.

% talk about GPU-based simulation before. Why GPU? what GPU-based tools before? NO REAL-TIME with realistic large system.
% we plan to use hybrid (GPU, CPU, DSP) platform to do real-time. What is our approach? what are the advantages of our approach?

To address the aforementioned issues associated with the existing primary fault detection algorithms, a hybrid approach is proposed. Sensor measurements, i.e., line currents and dc reactor voltages measured at local terminals, are first clustered and the best fault detector is selected among a detector pool by a learning algorithm. This detector pool consists of various existing detection algorithms, which exhibit different performances under different fault scenarios. The proposed hybrid primary detection algorithm provides the following advantages: i) all types of dc-side faults are detected, including pole-to-pole (P2P), pole-to-ground (P2G), and external dc fault; ii) detection is resilient to fault locations and impedances; iii) the detection method is robust to noisy sensor measurements.

The rest of this paper is organized as follows. The configuration of the test MTdc system is described in Section II. The proposed hybrid primary fault detection algorithm is introduced in Section III. In Section IV, the simulation results on the test MTdc system is provided and performance of the proposed method under different use cases is evaluated. The final conclusion is given in Section VI.

\section{The Test Multi-terminal HVdc System}

Fig. \ref{test_system} shows the layout of the test system adopted in this paper \cite{test_system1}. The test system, which represents a $\pm 320$ kV four-terminal meshed HVdc system, is comprised of four VSC stations connecting two offshore wind farms to two onshore ac systems. The transmission lines include Line12 and Line34 of 100 km length, Line13 and Line14 of 200 km length, and Line24 of 150 km length. Dc breakers are located at both ends of each HVdc link and between converters and buses. The detailed configuration of Line13 and Bus1 are depicted while other lines and buses use simplified representations in the figure. Further details of the test system along with its parameters are described in \cite{test_system1}.

\begin{figure}[t]
\begin{center}
\includegraphics[width=0.4\textwidth]{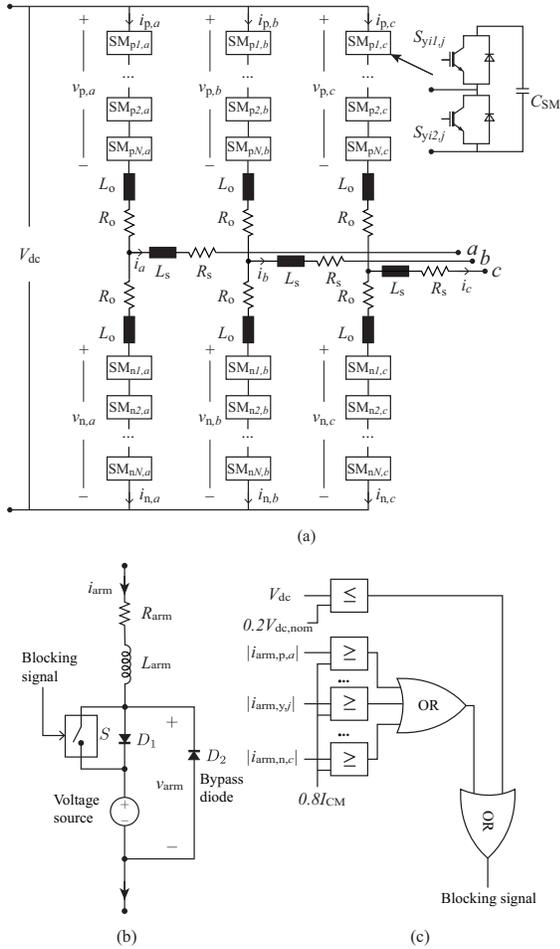}
\caption{Diagrams of the MMC models and their internal protection. a) circuit diagram and submodules (SMs); b) continuous equivalent MMC arm model with blocking/de-blocking capabilities \cite{test_system1,test_system2}; and c) MMC internal overcurrent and undervoltage protection.}
\label{internal_protection}
\end{center}
\vspace{-1\baselineskip}
\end{figure}

The neutral point of dc capacitors in the dc side of all VSCs are solidly grounded. The VSC stations, which are based on the well-known Modular Multilevel Converters (MMCs), shown in Fig. \ref{internal_protection}(a), are represented by their continuous equivalent models with blocking/de-blocking capabilities \cite{test_system1,test_system2}, as presented in Fig. \ref{internal_protection}(b). The blocking signals of IGBTs are triggered by the converter internal protection shown in Fig. \ref{internal_protection}(c), which consists of overcurrent and undervoltage protection. 
% The arm current threshold is set to be 80\% of the maximum instantaneous limit for the IGBT current, while the voltage threshold is selected to be 20\% of the nominal dc voltage. 
The cables are represented by the frequency-dependent model.

\begin{figure}[t]
\begin{center}
\includegraphics[width=0.4\textwidth]{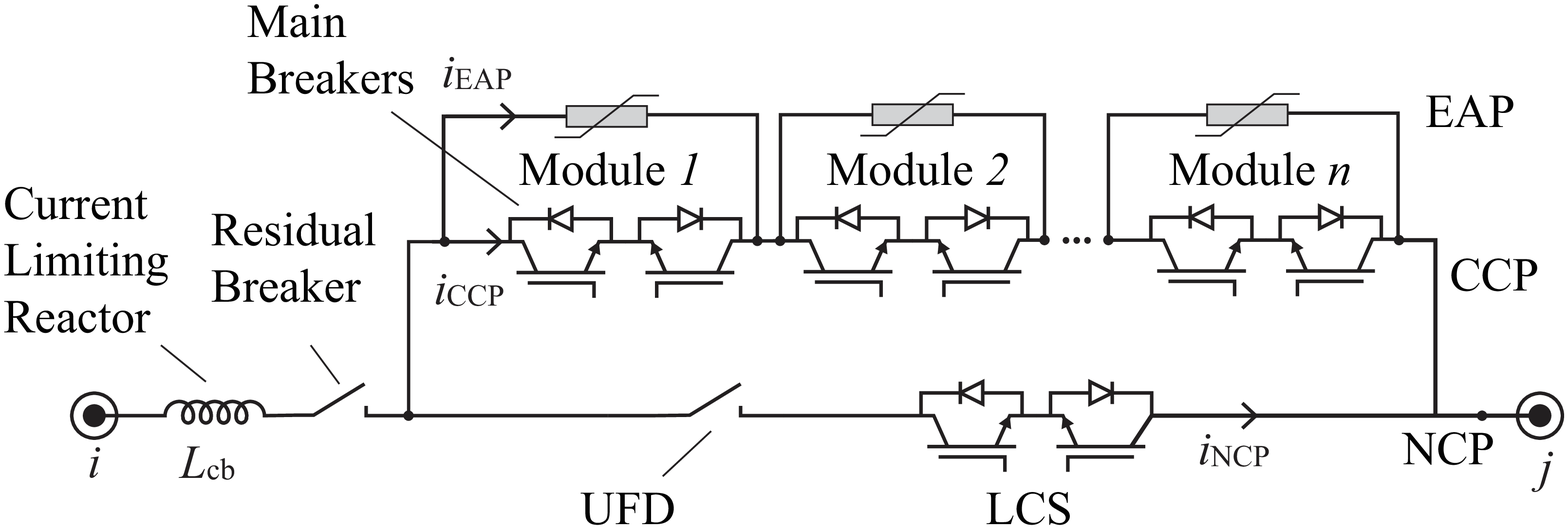}
\caption{Circuit diagram of the hybrid solid-state dc circuit breaker.}
\label{breaker}
\end{center}
\vspace{-1\baselineskip}
\end{figure}

The dc circuit breakers ($\text{CB}_{13P}$, etc.) used in the test system of Fig. \ref{test_system} are based on the hybrid solid-state dc circuit breaker \cite{breaker1,breaker4}, which breaks in a few milliseconds while offering low conduction
losses during normal operation. The detailed circuit diagram is presented in Fig. \ref{breaker}. Consisting of three paths, i.e., the nominal current path (NCP), the current commutation path (CCP), and the energy absorption path (EAP), a hybrid dc circuit breaker, is designed to clear a fault by commutating the fault current from the NCP to the CCP and EAP. During normal operation, the current flows through the ultra-fast disconnector (UFD) and the load commutation switch (LCS) in the NCP. Subsequent to a fault, the fault current is routed to the CCP, which is comprised of a number of identical modules with parallel connected main breakers and arrester banks. Once the current on the NCP reaches zero, the UFD opens immediately to prevent the LCS from exposure to high voltage. The main breakers are then tripped and the arrester banks are inserted simultaneously \cite{breaker1} or sequentially \cite{breaker3,breaker4} to extinguish the fault.

\section{Hybrid Primary Fault Detection Algorithm}

The proposed hybrid approach is built based upon the protection systems installed at the MTdc terminals. These protection systems aggregate the measurements collected from current and voltage sensors mounted on the points of interest. These measurements are processed by primary and backup fault detection algorithms to trip the corresponding dc circuit breakers when necessary.

\subsection{Layout of the Protection System}

\begin{figure}[t]
\begin{center}
\includegraphics[width=0.35\textwidth]{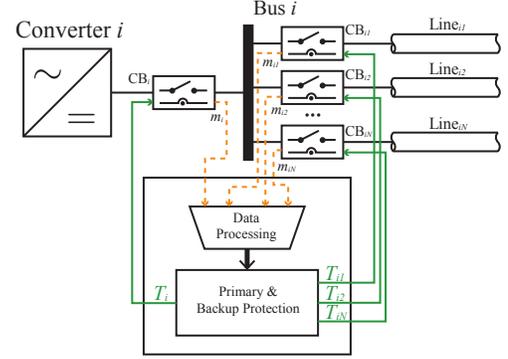}
\caption{The layout of the protection system at Bus $i$.}
\label{layout}
\end{center}
\vspace{-1\baselineskip}
\end{figure}

The layout of the proposed protection system is shown in Fig. \ref{layout}. For simplicity, the positive and negative lines are represented in one line view. As shown in Fig. \ref{layout}, Bus $i$ is connected with Converter $i$ through the breaker unit $\text{CB}_i$ and with other $N$ buses through breaker units $\text{CB}_{i1}, \text{CB}_{i2}, ..., \text{CB}_{iN}$. These breaker units consist of series connected dc circuit breakers and sensors that are placed on each circuit breaker and at the end of each HVdc link. The breakers are tripped by signals $T_i, T_{i1}, T_{i2}, ..., T_{iN}$, which are generated by their corresponding relaying algorithms in the primary and backup protection systems. The measurements $m_{i1}, m_{i2}, ..., m_{iN}$ consist of voltages across circuit breakers $v^{i1}, v^{i2}, ..., v^{iN}$, line currents $i^{lij}$, and terminal voltages beyond current limiting reactors $v^{lij}$ of those HVdc links that have one of their ends on the local Bus $i$, where $i,j$ are the two terminals of link $ij$. These measurements are captured with a sampling frequency $f_s$ and are then directly sent to the data processing unit. They serve as the input to both the primary and backup protection algorithms.

\subsection{Architecture of the Hybrid Primary Fault Detection Algorithm}

\begin{figure}[t]
\begin{center}
\includegraphics[width=0.4\textwidth]{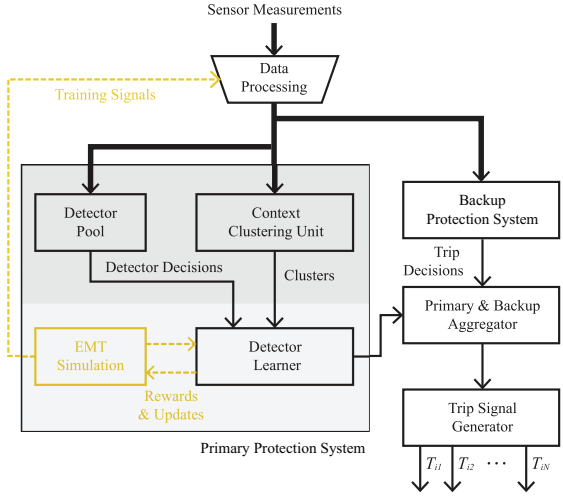}
\caption{Architecture of the proposed hybrid primary protection unit. The black lines and blocks represent the part of system implemented for online protection functionalities while the offline training components are depicted in yellow.}
\label{architecture}
\end{center}
\vspace{-1\baselineskip}
\end{figure}

The proposed hybrid primary fault detection algorithm employs a two-level architecture to extract information from the signal measurements. The detailed architecture is presented in Fig. \ref{architecture}. 

The sensor measurements are first associated to clusters capturing various modes of operation. Simultaneously, a pool of fault detectors is implemented. The detector pool consists of multiple candidate detectors that make decisions according to different subsets of measurements. The candidate detectors can be deployed based on the existing primary detection algorithms from literature \cite{detection1,detection7,detection9,detection8,detection3,detection4,detection2,detection6}.  For example, the threshold-based detectors \cite{detection1,detection7}, ROCOV-based detectors \cite{detection8,detection4}, and detectors based on QCD \cite{backup} are implemented in this study. However, the detector pool is fully customizable and modular and any detector can in principle be added to the pool.
% However, the input signals used in these detectors are not necessarily those used in the original work. 
% To cope with the bad data injected due to cyber intrusion, a bad data detector is implemented in the detector pool as well. 

The decisions made by the detector pool and the cluster identified by the context clustering unit then serve as the input to a detector learner making the final decision. The trip decisions made by the detector learner are aggregated with the decisions made by the backup protection. The trip signal generator finally assembles and applies the trip commands to the corresponding destination dc circuit breakers.

% The training of the detector learner is also illustrated by the dashed lines and yellow blocks in Fig. \ref{architecture}. While training the detector learner, it adaptively learns from previous experience generated by the training simulator. The simulator provides the feedback and rewards to the detector learner to help update its internal parameters. This online learning is proceeded in an iterative manner. A new training episode is restarted and new sets of simulated signals are sent to the data processing unit within each iteration. This online learning does not require a large number of training iterations, which makes it suitable to be applied in the simulation-based setting. 

As illustrated by the dashed lines and yellow blocks in \mbox{Fig. \ref{architecture}}, the detection learner is trained offline with samples generated by an electromagnetic transients (EMT) simulation. The simulation provides feedback and rewards to the detector learner to help update its internal parameters. These parameters are used to generate the final decision based on the decisions made by individual detectors from the pool.
% This online learning is proceeded in an iterative manner. A new training episode is restarted and new sets of simulated signals are sent to the data processing unit within each iteration. This online learning does not require a large number of training iterations, which makes it suitable to be applied in the simulation-based setting. 

\subsection{Context Clustering Unit}

\begin{figure}[t]
\begin{center}
\includegraphics[width=0.35\textwidth]{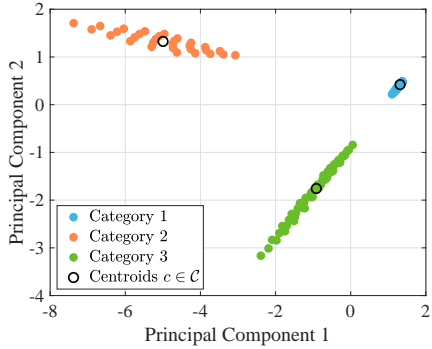}
\caption{Context clustering using observations from Terminal 1 and Line13.}
\label{kmeans}
\end{center}
\vspace{-1\baselineskip}
\end{figure}

The sensor measurements are first clustered by the context clustering unit. The idea of clustering originates from the observation that different detectors perform differently under various operation conditions, which are separable given the sensor measurements, as seen in Fig. \ref{kmeans} later. These separate operation conditions form a ``context" for the detectors. Upon identifying a context, detectors in the pool are assigned different weights.
% , which indicates their fatalities. 
This context clustering helps the system grasp a better understanding of the signals by providing the context to which they belong. 

The $k$-means clustering \cite{MLbook} is applied for context clustering. The $k$-means clustering algorithm strives to partition the given $d$-dimensional ($d=6$ in this work) observations into $k$ clusters with a minimized within-cluster variance. The details of this standard method can be found in \cite{MLbook} and are not repeated here.

% Six-dimensional observations are used as the training sets for the clustering algorithm. 
As shown in Fig. \ref{layout}, there are $N$ HVdc links connected with the $i$-th MTdc terminal. For each connected HVdc link $j$, the line currents $i^{lij}$, line voltages beyond current limiting reactors $v^{lij}$, and voltages across the dc circuit breakers $v^{ij}$ from both positive and negative poles are collected. The six measurements form six dimensional feature vectors. The dataset is generated from simulations of P2P, low impedance P2G, and high-impedance P2G faults located at \SI{10}{\kilo\meter} intervals from each HVdc link. The fault resistances are also varied from \SI{1}{\ohm} to \SI{500}{\ohm}.

% The Gaussian Mixture Models (GMM) is applied in this work for the context clustering.

% A GMM is a probabilistic model that takes a mixture of uni-modal distributions, i.e., Gaussian distributions \cite{MLbook}. It attempts to model the input measurements by looking for the best mixture of multi-dimensional Gaussian probability distributions. Consider the probability model of a mixture of K Gaussians,

% \begin{equation}
% p(X) = \sum_{k=1}^{K} \pi_{k} \cdot \mathcal{N}(X|\mu_{k}, \Sigma_{k}),
% \end{equation}
% where $X$ is the aggregation of input observed measurements samples. $\pi_{k}$, $\mu_{k}$, and $\Sigma_{k}$ are the mixing proportion, mean and variance of $k$th distribution, respectively. $\mathcal{N}(X|\mu_{k}, \Sigma_{k})$ is the probability density of a single Gaussian mixture component, written as

% \begin{equation}
% \mathcal{N}(X|\mu_{k}, \Sigma_{k}) = \frac{1}{|\Sigma_{k}|^{\frac{1}{2}}(2\pi)^{\frac{d}{2}}} \cdot e^{-\frac{1}{2}(X-\mu_{k})^T \Sigma_{k}^{-1} (X-\mu_{k})}.
% \end{equation}

% The expectation–maximization (EM) algorithm is used to learn the parameters, i.e., $\pi_{k}$, $\mu_{k}$, and $\Sigma_{k}$ of the GMM. The details of the EM can be found in \cite{MLbook} and is not repeated here.

The output of the context clustering unit is a cluster label $c \in \mathcal{C}$, where $\mathcal{C}$ is a finite set for context cluster labels. This label indicates which cluster the given sensor measurements belong to. 
% This clustering unit also greatly improves the performance of the overall primary detection with the ability to 
% distinguish bad or manipulated data from normal sensor measurements. It also helps 
% separate external faults from internal dc faults. 
This cluster label is sent as the input to the lower-level hierarchy of the system.

On each MTdc terminal, independent clusters in context clustering unit are trained for each line/cable. The size of the datasets used for training varies given the length of the target line/cable. For example, 271 samples are used for training the cluster for Line13 on Terminal 1. These samples consist of 35 P2P, 70 low-impedance P2G, 70 high-impedance P2G, and 96 normal operation cases.

It is also critical to determine the hyper-parameter $k$ in the $k$ means algorithm. Given the number of different fault types considered to generate the data, 2, 3 and 4 are potential candidates for the parameter $k$. These three candidates are evaluated through calculating their silhouette values \cite{MLbook}, which are 0.749, 0.918, and 0.872, respectively. A higher silhouette value indicates a better matching of samples to their own cluster rather than the neighboring clusters. Therefore, $k=3$ is the value used in this study.

The clustering results are presented in Fig. \ref{kmeans}. The training simulation samples are generated at different locations on Line13 using various settings, as described above. The six-dimensional training observations are collected at Terminal 1 and used for the $k$-means clustering, where $k=3$. To better visualize the results, principal component analysis (PCA) \cite{MLbook} is performed on the six-dimensional data. The top two principal components are plotted in Fig. \ref{kmeans}. Three categories, labelled 1 to 3, are separated apart. These categories provide an idea of the fault context of faults on Line13 observed from terminal 1. For example, category 1, located in the top right corner of the graph, aggregates the observations from high impedance P2G faults and normal cases. The observations are not separable on all dimensions of the input space because both current and voltages measurements under high-impedance P2G faults do not clearly deviate from normal operation, especially when there is noise in some of the measurements. The clustering is not intended to precisely detect or classify faults, instead, it is used to provide a coarse description of what the signals look like. These clusters then help the detectors from the detector pool to make improved decisions.

\subsection{Detector Pool}

The sensor measurements are sent to a pool of $N$ detectors in parallel. Each detector within the detector pool takes a subset of the measurements as its input and makes decisions. 

As summarized in the literature review, none of the existing fault detectors is perfect. Different detectors work best under different cases. Multiple existing detectors are implemented in the detector pool, including the threshold-based detectors \cite{detection1,detection7}, ROCOV-based detectors \cite{detection8,detection4}, and QCD-based detectors \cite{backup}. These detectors make independent decisions $d_n, n \in N$, which are then fed to the detector learner.

\subsection{Detector Learner}

The detector learner attempts to combine the decisions made by the context clustering unit and the detector pool, and is inspired by ideas from ensemble learning \cite{MLbook}. The detector pool is designed to make decisions using the inputs from multiple less powerful detectors in the detector pool. The final decision made by the ensemble method performs better than what could be obtained by any single detector alone. 

For each cluster label $c$ obtained from the context clustering unit, one detector learner is trained. Given the independent decisions $d_{n,c}$ made by each detector $n$ under each cluster label $c$, the final decision is made by taking a weighted majority over these inputs. 

Specifically, for each cluster $c \in \mathcal{C}$, a weight $w_{n,c}, n \in \{1...N\}$ is assigned to each detector. The weights from all detectors satisfy $\sum_{n=1}^{N} w_{n,c} = 1$. Each detector makes a binary decision to address a fault, and an accumulated metric $h_c$ is calculated as 
\begin{equation}
h_c = \sum_{n=1}^{N} w_{n,c} \times d_{n,c}, c \in \mathcal{C}.
\end{equation}

The final decision $d_c$ made by the detector learner is 1 if $h_c$ is higher than 0.5 and 0 otherwise. It may be noted that the weights learned during training are typically different for different clusters. Detectors that make better decision and provide more robust performance tend to be assigned higher weights. The weights are learned through a single-round of training, as shown in Algorithm \ref{alg:Exp3}.

\SetEndCharOfAlgoLine{}
\begin{algorithm}[ht]
\SetAlgoNoLine
\KwIn{Training samples $\boldsymbol{x_{i}}$ from EMT simulation} 
Initialize $w_{n,c} = 1/N$ \\
\For{\normalfont{each cluster label } $c \in \mathcal{C}$}{
    Filter out the $m_c$ training samples matching label $c$ \\
    Initialize counter $\text{cnt}_{n} = 0$ for detector $n$ \\
    \For{\normalfont{each detector } $n = 1,...,N$}{
        Test each training sample and accumulate decision $d_n$ to the counter $\text{cnt}_{n}$\\
    }
    Calculate correct rate $r_{n,c} = \frac{\text{cnt}_{n}}{m_c}$ for each detector \\
    Update the weights as $w_{n,c} = \frac{r_{n,c}}{\sum_{n=1}^{N}r_{n,c}}$ \\
}
\caption{Weighted majority update algorithm}
\label{alg:Exp3}
\end{algorithm}

Initially, all weights are set to $w_{n,c} = 1/N$, and the training samples $\boldsymbol{x_{i}}$ are assigned cluster labels $c$ by the context cluster. For each cluster label, the samples with same label are filtered out. The number of samples for each cluster label is denoted as $m_c$. The detectors are then tested on these samples by accumulating their binary decisions into counters. Finally, new weights $w_{n,c}$ for detectors are computed and normalized proportional to their correct rate.

% Given each context cluster label $c$, an instance of Exponential-weight algorithm for Exploration and Exploitation (Exp3) algorithm \cite{RLbook} is applied to learn a policy $\tau$, which maps from a context label $c$ to a detector in the detector pool. The Exp3 algorithm strives to minimize the regret defined by

% \begin{equation}
% R_T = \min_{\tau: C \rightarrow \{1,...,N\}} \mathbb{E} \Big[ \sum_{t=1}^{T} l_{d_t, t} - \sum_{t=1}^{T} l_{\tau(c_t), t} \Big],
% \end{equation}
% where $c_t$ denotes the context label at time instance $t$. $l_{d_t, t}$ and $l_{\tau(c_t), t}$ are the loss incurred by the detector learner and the loss incurred by the best mapping from context label to detector, respectively. 

% \SetEndCharOfAlgoLine{}
% \begin{algorithm}[ht]
% \SetAlgoNoLine
% \KwIn{$\lambda$, uniformly initialized $p_1^{(c)}$}
% \For{\normalfont{each time instance } $t = 1,...,T$}{
%     Collect context label $c_t \in C$ \\
%     Draw a decision $d_t$ from distribution $p_{i,t}^{(c_t)}$ \\
%     Compute estimated loss for each detector \\
%     Update the distribution $p_{i,t+1}^{(c_t)}$
% }
% \caption{Exp3 for training detector learner}
% \label{alg:Exp3}
% \end{algorithm}

% The Exp3 algorithm is presented in Algorithm \ref{alg:Exp3}. $p_{i,t}^{(c_t)}$ denotes the probability of choosing detector $i = 1,...,N$ under context label $c_t$ at time $t$. The detector learner is trained online, as illustrated in Fig. \ref{architecture}. The final decision made by the detector learner is the best decision from the detector pool under the given context.  

\section{Study Results}

Study results are now presented to evaluate the performance and effectiveness of the proposed hybrid primary fault detection algorithm under four scenarios: a) a P2P fault under normal operation conditions; b) a low-impedance P2G fault; and c) a high-impedance P2G fault.
% , and d) signals manipulated by cyber intrusion. 
The system \cite{test_system1} is modeled in the PSCAD/EMTDC software environment with its parameters listed in Table \ref{parameters}. A sampling frequency of $f_s = \SI{50}{\kilo\hertz}$ is adopted in all simulations.

\begin{table}[ht]
\centering
\caption{Converter and system parameters \cite{test_system1}}
\begin{tabular}{ccc}
\hline
 & Conv. 1,2,3 & Conv. 4
 \\\hline
 Rated power [MVA] & 900 & 1200 \\
 Ac grid voltage [kV] & 400 & 400 \\
 Dc grid voltage [kV] & $\pm$320 & $\pm$320 \\
%  Transformer $u_k$ [pu] & 0.15 & 0.15 \\
%  Ac grid reactance $X_{\text{ac}}$ [$\Omega$] & 17.7 & 13.4 \\
%  Ac grid resistance $R_{\text{ac}}$ [$\Omega$] & 1.77 & 1.34 \\
 \hline
 Arm capacitance $C_{\text{arm}}$ [$\mu F$] & 29.3 & 39 \\
 Arm reactor $L_{\text{arm}}$ [mH] & 84.8 & 63.6 \\
 Arm resistance $R_{\text{arm}}$ [$\Omega$] & 0.885 & 0.67 \\
 Bus filter reactor [mH] & 50 & 50 \\
\hline
\end{tabular}
\label{parameters}
\vspace{-1\baselineskip}
\end{table}

\subsection{The Implemented Detector Pool and Detector Learner}

There are four detectors, indexed from 1 to 4, implemented in the detector pool, i.e., current threshold, current derivative, ROCOV, and QCD based detector, respectively.

The current threshold-based detector takes the line current as the input and simply compares it with a pre-determined threshold. The threshold is selected to be 1.25 times the nominal line current in the study to balance the detection speed and noise withstand capability.

The current derivative and ROCOV-based detectors \cite{detection7,detection8} compute the derivative of line currents and voltages beyond current limiting reactors, respectively. To make these detectors more robust under noisy signals, a moving average filtering is performed on the given inputs. A higher window size works better on filtering but slows down the detector. Considering this trade-off, the window size is selected to be three in this work. 

The QCD-based detector \cite{backup} takes the same inputs as the ROCOV-based detector. However, the sensor measurements are not pre-processed using the moving average filter.

As shown in Fig. \ref{kmeans}, there are three cluster labels. The updated weights $w_{n,c}$ for each detector obtained after training given the cluster labels are presented in Table \ref{weights}.

\begin{table}[ht]
\centering
\caption{The weights used in detector learner}
\begin{tabular}{ccccc}
\hline
 Detector & Current threshold & Current derivative & ROCOV & QCD
 \\\hline
Label 1 & 0.365 & 0.044 & 0.080 & 0.511 \\
Label 2 & 0.193 & 0.281 & 0.281 & 0.246 \\
Label 3 & 0.246 & 0.262 & 0.262 & 0.230 \\
\hline
\end{tabular}
\label{weights}
\vspace{-1\baselineskip}
\end{table}

\subsection{P2P Fault}

In this scenario, the system of Fig. \ref{test_system} is subjected to a P2P fault on Line13, which is \SI{105}{\kilo\meter} away from Terminal 1. This fault is triggered at $t = \SI{0.71}{\second}$. The propagation of travelling wave takes around \SI{0.5}{\milli\second} to reach this terminal \cite{breaker_selection}. 

\begin{figure*}[t]
\begin{center}
\includegraphics[width=0.99\textwidth]{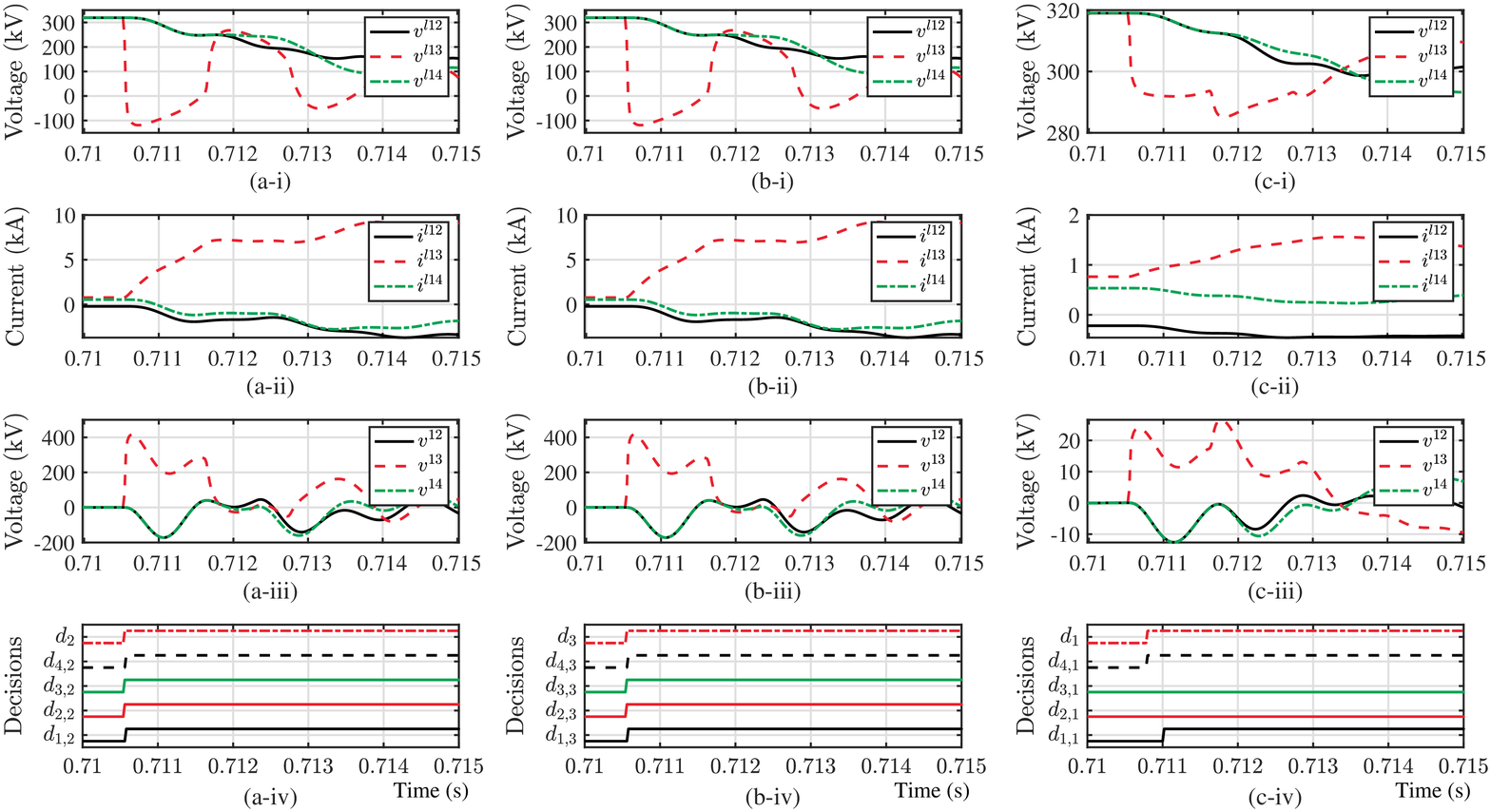}
\caption{Simulation results with P2P, low-impedance P2G, and high-impedance P2G faults on Line13 located \SI{105}{\kilo\meter} away from Terminal 1: (a-i) line voltages measured at Converter 1, (a-ii) line currents measured at Converter 1, (a-iii) voltages across dc reactors close to Converter 1, and (a-iv) decisions made by the detectors from detector pool and detector learner during a P2P fault; (b-i) line voltages measured at Converter 1, (b-ii) line currents measured at Converter 1, (b-iii) voltages across dc reactors close to Converter 1, and (b-iv) decisions made by the detectors from detector pool and detector learner during a low impedance P2G fault; (c-i) line voltages measured at Converter 1, (c-ii) line currents measured at Converter 1, (c-iii) voltages across dc reactors close to Converter 1, and (c-iv) decisions made by the detectors from detector pool and detector learner during a high-impedance P2G fault.}
\label{normal_case_all}
\end{center}
\vspace{-1\baselineskip}
\end{figure*}

The simulation results are demonstrated in Figs. \ref{normal_case_all}(a-i)-(a-iv). The line voltages measured beyond the current limiting reactors $v^{lij}$, line currents $i^{lij}$, and voltage across the dc reactors $v^{ij}$ are presented in Figs. \ref{normal_case_all}(a-i)-(a-iii), respectively. The measurements from negative poles are also used in the proposed approach but are not shown in the figures. In the detector pool, four types of detectors are implemented based on these measurements. The ROCOV- and QCD-based detectors are applied with signals $v^{lij}$. Signal $i^{lij}$ are used in current threshold and current derivative based detectors. 

Fig. \ref{normal_case_all}(a-iv) shows the decisions made by the four detectors $d_{n,2}$ and the final decision $d_{2}$. The context is clustered to be label 2. Detector 1 (current threshold) and Detector 4 (QCD) are slightly slower than the other two derivative-based detectors (2 and 3). According to Table \ref{weights}, the summation of weights from detectors 2 and 3 are higher than 0.5, the final decision is made when these two detectors are triggered. This makes the proposed algorithm operate as fast as the fastest detectors. The trip signal of dc circuit breaker $\text{CB}_{13}$, $T_{13}$ is triggered at $t = \SI{0.7111}{\second}$, less than \SI{1}{\milli\second} after the arrival of fault wave front. The rest of dc circuit breakers are safely kept untripped. 

% Fig. \ref{p2p}(d) shows the final trip signals generated by the detector learning. The trip signal of dc circuit breaker $\text{CB}_{13}$, $T_{13}$ is triggered at $t = \SI{0.7111}{\second}$, less than \SI{1}{\milli\second} after the arrival of fault wave front. The rest of dc circuit breakers are safely kept untripped. 

% \begin{figure}[t]
% \vspace{-1\baselineskip}
% \centering
% \begin{subfigure}{\columnwidth} 
% \includegraphics[width=\textwidth]{p2p_a.eps}
% \label{p2p_a}
% \vspace{-1\baselineskip}
% \end{subfigure}
% \begin{subfigure}{\columnwidth} 
% \includegraphics[width=\textwidth]{p2p_b.eps}
% \label{p2p_b}
% \vspace{-1\baselineskip}
% \end{subfigure}
% \begin{subfigure}{\columnwidth} 
% \includegraphics[width=\textwidth]{p2p_c.eps}
% \label{p2p_c}
% \vspace{-1\baselineskip}
% \end{subfigure}
% \begin{subfigure}{\columnwidth} 
% \includegraphics[width=\textwidth]{p2p_d.eps}
% \label{p2p_d}
% \vspace{-1\baselineskip}
% \end{subfigure}
% \caption{Simulation results with a P2P fault in the middle of Line13: (a) line voltages measured at Converter 1, (b) line currents measured at Converter 1, (c) voltages across dc reactors close to Converter 1, and (d) decisions made by the detectors from detector pool and detector learner.}
% \label{p2p}
% \end{figure}

\subsection{P2G Fault}

Another fault type that has been tested is the low-impedance P2G fault. The positive pole of Line13 is grounded at \SI{105}{\kilo\meter} away from Terminal 1. The measured signals are presented in Figs. \ref{normal_case_all}(b-i)-(b-iv).

% \begin{figure}[t]
% \vspace{-1\baselineskip}
% \centering
% \begin{subfigure}{\columnwidth} 
% \includegraphics[width=\textwidth]{p2g_a.eps}
% \label{p2g_a}
% \vspace{-1\baselineskip}
% \end{subfigure}
% \begin{subfigure}{\columnwidth} 
% \includegraphics[width=\textwidth]{p2g_b.eps}
% \label{p2g_b}
% \vspace{-1\baselineskip}
% \end{subfigure}
% \begin{subfigure}{\columnwidth} 
% \includegraphics[width=\textwidth]{p2g_c.eps}
% \label{p2g_c}
% \vspace{-1\baselineskip}
% \end{subfigure}
% \begin{subfigure}{\columnwidth} 
% \includegraphics[width=\textwidth]{p2g_d.eps}
% \label{p2g_d}
% \vspace{-1\baselineskip}
% \end{subfigure}
% \caption{Simulation results with a P2G fault in the middle of Line13: (a) line voltages measured at Converter 1, (b) line currents measured at Converter 1, (c) voltages across dc reactors close to Converter 1, and (d) decisions made by the detectors from detector pool and detector learner.}
% \label{p2g}
% \end{figure}

Since only the measurements from the positive poles are shown in the figures, the waveforms look almost identical to the P2P case. The signals from negative poles, however, do not significantly deviate from their nominal states. This P2G fault is detected by the proposed primary detector and the trip signal for dc circuit breaker $\text{CB}_{13}$ is generated.

\subsection{High-Impedance P2G Fault}

High-impedance P2G faults are hard to detect since signals deviate little from nominal values in this scenario. The context of this case is clustered to be label 1.

% \begin{figure}[t]
% \vspace{-1\baselineskip}
% \centering
% \begin{subfigure}{\columnwidth} 
% \includegraphics[width=\textwidth]{p2g_hi_a.eps}
% \label{p2g_hi_a}
% \vspace{-1\baselineskip}
% \end{subfigure}
% \begin{subfigure}{\columnwidth} 
% \includegraphics[width=\textwidth]{p2g_hi_b.eps}
% \label{p2g_hi_b}
% \vspace{-1\baselineskip}
% \end{subfigure}
% \begin{subfigure}{\columnwidth} 
% \includegraphics[width=\textwidth]{p2g_hi_c.eps}
% \label{p2g_hi_c}
% \vspace{-1\baselineskip}
% \end{subfigure}
% \begin{subfigure}{\columnwidth} 
% \includegraphics[width=\textwidth]{p2g_hi_d.eps}
% \label{p2g_hi_d}
% \vspace{-1\baselineskip}
% \end{subfigure}
% \caption{Simulation results with a high impedance P2G fault in the middle of Line13: (a) line voltages measured at Converter 1, (b) line currents measured at Converter 1, (c) voltages across dc reactors close to Converter 1, and (d) decisions made by the detectors from detector pool and detector learner.}
% \label{p2g_hi}
% \end{figure}

In this scenario, a high-impedance P2G fault is imposed on the positive pole of Line13 (\SI{105}{\kilo\meter} from Bus 1). A \SI{300}{\ohm} fault impedance is inserted between the fault location and the ground. With a higher fault impedance applied, the drop of voltage magnitude is even smaller compared to the low-impedance P2G case. 

The performance of four candidate detectors varies significantly in this scenario. The parameters of detectors 2 and 3 are selected to be not too sensitive to the change of current or voltage derivatives to withstand noisy signals. Therefore, under this scenario, they are not able to detect the fault. Detector 1 successfully detects the fault. However, it is slow since it has to wait for the fault current to go across the pre-defined threshold. The rate of rise of current is not as high as the P2P or low impedance P2G cases. Therefore, it takes longer for detector 1 to be triggered. Detector 4, however, performs best in this scenario. It is not as fast as in previous cases due to the smaller deviation of voltage signals, but it is faster than detector 1. According to Table \ref{weights}, detector 4 is assigned the highest weight by the detector learner, which means that it is the most trustworthy detector when the context is labelled 1. The final decision is made and the trip signal for dc circuit breaker $\text{CB}_{13}$ is generated when detector 4 is triggered, as shown in Fig. \ref{normal_case_all}(c-iv).

\begin{figure}[t]
\begin{center}
\includegraphics[width=0.42\textwidth]{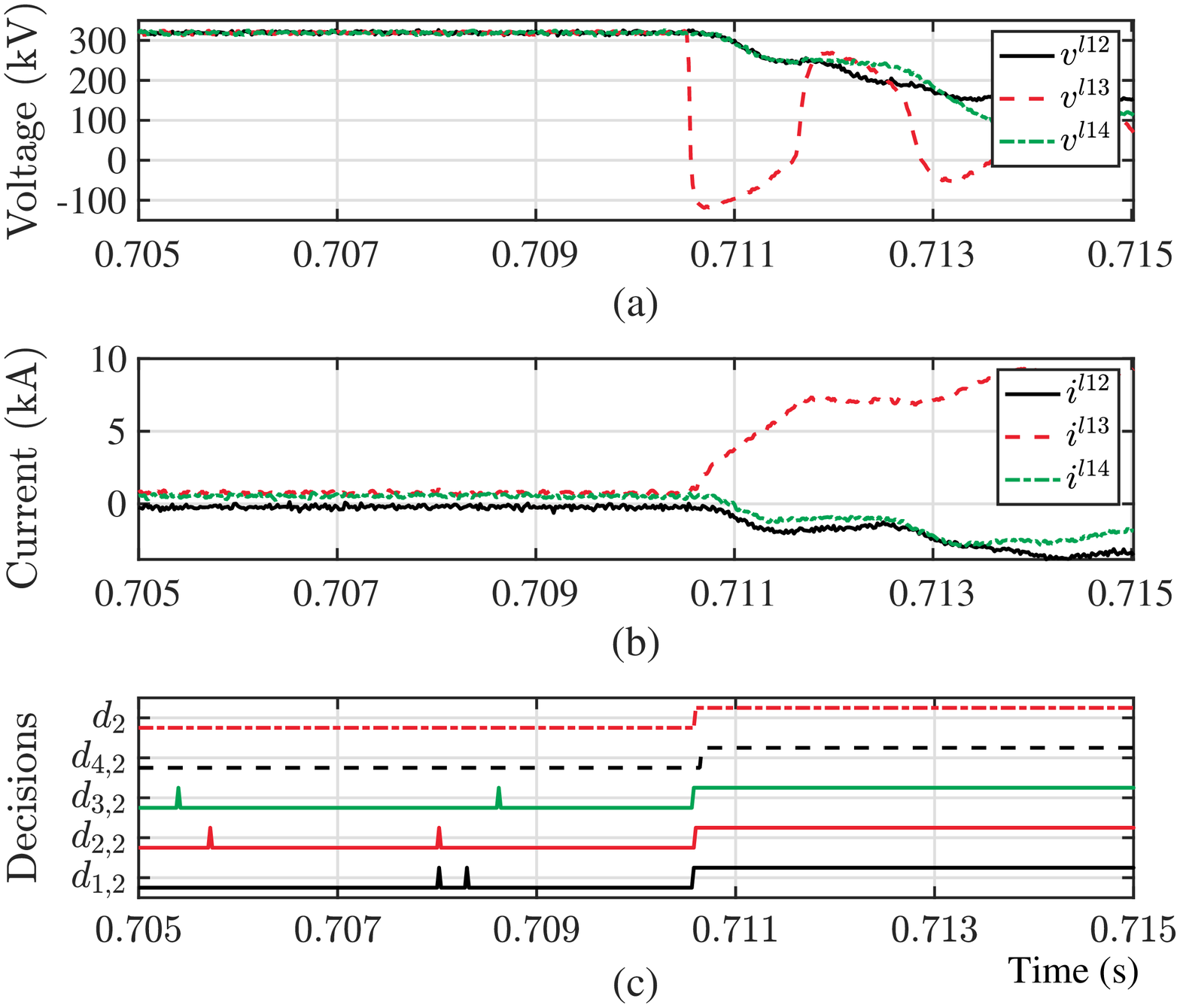}
\caption{Simulation results with noisy sensor measurements under a P2P fault on Line13 located \SI{105}{\kilo\meter} away from Terminal 1: (a) line voltages measured at Converter 1, (b) line currents measured at Converter 1, and (c) decisions made by the detectors from detector pool and detector learner.}
\label{noise_cases}
\end{center}
\vspace{-1\baselineskip}
\end{figure}

\subsection{Noisy Signals}

To evaluate the robustness of the proposed hybrid detector, the primary protection unit is tested with sensor measurements contaminated by Gaussian noise, as shown in Figs. \ref{noise_cases}(a)-(c).

As illustrated in Fig. \ref{noise_cases}(c), individual candidate detectors in the detector pool may make wrong decisions prior to the fault occurrence. False alarms can be triggered if the final tripping decision is made using only one candidate detector. This noise vulnerability of single detector can be mitigated by allowing tripping decision only after observing consistent alarms over a longer period. However, such operation will slow down the fault detection. The proposed hybrid detector, which relies on multiple detectors in the pool to make its final decision, improves the robustness to noisy sensor measurements without sacrificing the detection speed.

\subsection{False Alarms and Detection Failure}

\begin{figure}[t]
\begin{center}
\includegraphics[width=0.3\textwidth]{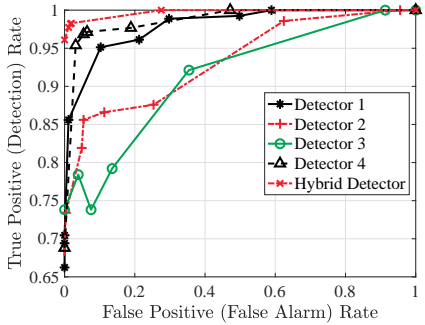}
\caption{ROC curves of the four candidate detectors in the detector pool and the proposed hybrid detector.}
\label{ROC}
\end{center}
\vspace{-1\baselineskip}
\end{figure}

A plot of the Receiver Operating Characteristics (ROC) curves \cite{MLbook} is presented in Fig. \ref{ROC}, which shows the performance of different detectors under various threshold settings. The true positive (detection) rate represents the capability of a detector to capture the faults, while the false positive rate denotes the possibility of triggering false alarm from the detector. In ROC standard analysis, a higher Area Under The Curve (AUC) indicates a higher performance. As shown in Fig. \ref{ROC}, the proposed hybrid detector achieves best performance compared with the four individual detectors.

The proposed primary fault detection algorithm improves robustness and reliability by performing a majority voting from detectors in the detector pool. As a result, the overall performance metrics such as miss-detection and false alarm rates are as good as the performance metrics of the best detector in the pool. The current threshold-based detector is fast in P2P and low-impedance P2G fault cases but its detection speed is significantly reduced in fault scenarios where fault current increases slowly. The QCD detector is robust to signals where noise and spikes are introduced, at the expense of sacrificing the detection speed. The current derivative and ROCOV based detectors can be tuned to be sensitive to high impedance faults, but they will be more likely to raise false alarms. These trade-offs worsen the performance of individual detector and are hard to balance. The proposed hybrid approach, however, combines the best features of each detector and avoid their drawbacks. The candidate detectors can thus be tuned to work best on particular scenarios, and are not necessarily well balanced to all cases. The choices and optimized tuning of the detector pool will be addressed in future work.

\section{Conclusion}

In this paper, a hybrid primary fault detection algorithm is proposed for the MTdc systems. Instead of relying on a single detector to detect the fault, the proposed algorithm makes its tripping decision using a learning algorithm which considers all the decisions made by multiple candidate detectors among a detector pool. This detector pool consists of various existing detection algorithms, each performing differently across fault scenarios. The proposed hybrid primary detection algorithm offers the following advantages: i) all types of dc-side faults are covered, including pole-to-pole (P2P), pole-to-ground (P2G), and external dc faults; ii) various fault locations and impedances are covered; iii) the system is robust to noisy sensor measurements. The proposed algorithm is fully modular and can be extended and improved by using different set of candidate detectors to cover challenging fault scenarios like cyber intrusions in future development. 

% Performance and effectiveness of the proposed algorithm are evaluated and verified based on time-domain simulations in the PSCAD/EMTDC environment. The results confirm satisfactory operation, accuracy, and speed of the proposed algorithms under various fault scenarios.

% In this paper, a local measurement-based backup protection algorithm for MTdc systems is proposed. The proposed algorithm that is based on the quickest change detection (QCD) technique, achieves fast and accurate backup protection functionality for the primary relay to ensure a higher reliability in the system. The proposed method can be readily extended to different grid configurations and is able to cooperate with different primary protection algorithms and breaker configurations. Performance and effectiveness of the proposed algorithm are evaluated and verified based on time-domain simulation studies in the PSCAD/EMTDC environment. The results confirm satisfactory performance of the proposed algorithm in terms of accuracy, robustness, and speed under various fault scenarios.

\ifCLASSOPTIONcaptionsoff
  \newpage
\fi

% \balance
\bibliographystyle{IEEEtran}
\bibliography{IEEEabrv,reference}

% \appendices
% \numberwithin{equation}{section}
% \renewcommand{\theequation}{\thesection.\arabic{equation}}

% \section{DAEs describing generator dynamics} \label{appen_a}

% \begin{align}
%     \frac{1}{\omega_s} \frac{d \psi_d}{dt} &= R_s I_d + \frac{\omega}{\omega_s}\psi_q + V_d \label{eq:gen1}\\
%     \frac{1}{\omega_s} \frac{d \psi_q}{dt} &= R_s I_q - \frac{\omega}{\omega_s}\psi_d + V_q \label{eq:gen2}\\
%     \frac{1}{\omega_s} \frac{d \psi_0}{dt} &= R_s I_0 + V_0 \label{eq:gen3}\\
%     T_{d0}' \frac{d E_q'}{dt} &= -E_q' - (X_d - X_d')[I_d - \frac{X_d'-X_d''}{(X_d'-X_{ls})^2}(\psi_{1d} \nonumber\\
%     & + (X_d' - X_{ls})I_d - E_q')] + E_{fd} \label{eq:gen4}\\
%     T_{d0}'' \frac{d \psi_{1d}}{dt} &= -\psi_{1d} + E_q' - (X_d' - X_{ls})I_d \label{eq:gen5}\\
%     T_{q0}' \frac{d E_d'}{dt} &= -E_d' - (X_d - X_d')[I_d - \frac{X_d'-X_d''}{(X_d'-X_{ls})^2}(\psi_{2q} \nonumber\\
%     & + (X_q' - X_{ls})I_q + E_d')] \label{eq:gen6}\\
%     T_{q0}'' \frac{d \psi_{2q}}{dt} &= -\psi_{2q} + E_d' - (X_q' - X_{ls})I_q \label{eq:gen7}\\
%     \frac{d \delta}{dt} &= \omega - \omega_s \label{eq:gen8}\\
%     \frac{2H}{\omega_s} \frac{d\omega}{dt} &= T_M - (\psi_d I_q - \psi_q I_d) - T_{FW} \label{eq:gen9}\\
%     \psi_d &= -X_d'' I_d + \frac{X_d''-X_{ls}}{X_d'-X_{ls}}E_q' + \frac{x_d'-X_d''}{X_d'-X_{ls}}\psi_{1d} \label{eq:gen10}\\
%     \psi_q &= -X_q'' I_q + \frac{X_q''-X_{ls}}{X_q'-X_{ls}}E_d' + \frac{x_q'-X_q''}{X_q'-X_{ls}}\psi_{2q} \label{eq:gen11}\\
%     \psi_0 &= -X_{ls} I_0 \label{eq:gen12}
% \end{align}

\end{document}